\documentclass[runningheads]{llncs}

\usepackage{graphicx}
\usepackage{amsmath}
\usepackage{amssymb}
\usepackage{booktabs}
\usepackage{multirow}
\usepackage{hyperref}

\begin{document}

\title{Native-Space 3D CarveMix for Multi-Site T1w Stroke Segmentation}

\author{Dexter Wen Jie Teo \inst{1,2} \and Kumaradevan Punithakumar \inst{2}}
\authorrunning{Teo and Punithakumar}
\institute{College of Computing and Data Science, Nanyang Technological University, Singapore \and
Department of Radiology and Diagnostic Imaging, University of Alberta, Edmonton, Canada \\
Corresponding author: \email{WTEO030@e.ntu.edu.sg}}

\maketitle

\begin{abstract}
Segmenting ischemic stroke lesions on T1-weighted (T1w) MRI acquired across different scanners and protocols without intensity standardization is difficult because lesions are subtle and share intensity characteristics with cerebrospinal fluid. Standard deep learning architectures trained across multiple centers plateau around Dice 0.66, with acute lesions ($\le 7$ days post-stroke) performing substantially worse due to severe sample scarcity. We combine a MedNeXt-L ($k=5$) backbone with on-the-fly 3D CarveMix augmentation that pastes real lesion patches into recipient training patches. By generating synthetic lesion placements dynamically within each fold with subject-level split isolation, the model sees more diverse lesion patterns without requiring pre-generated copies on disk. We evaluate on 1,453 native T1w scans from 55 clinical centers in the ISLES 2026 challenge. MedNeXt-L with CarveMix achieved a mean five-fold cross-validation Dice of 0.648 after 500 epochs. For comparison, the estimated Dice for plain MedNeXt-L at epoch 500 was 0.630, obtained by interpolation from its 1,000-epoch training trajectory.

\keywords{Ischemic Stroke Segmentation \and T1-Weighted MRI \and ISLES 2026 \and 3D CarveMix \and MedNeXt.}
\end{abstract}

\section{Introduction}

Stroke remains one of the leading causes of death and long-term disability worldwide \cite{feigin2021gbd}. Precise delineation of stroke lesion volume and anatomical location is critical for predicting long-term functional outcomes and evaluating therapeutic efficacy in clinical trials \cite{liew2022atlas}. In acute clinical settings, stroke diagnosis relies on multi-parametric MRI protocols, specifically Diffusion-Weighted Imaging (DWI) and Fluid-Attenuated Inversion Recovery (FLAIR), which provide high contrast for acute stroke lesions \cite{petzsche2022isles}. However, in large observational cohorts and retrospective neuroimaging datasets, DWI or FLAIR sequences are often unavailable \cite{liew2022atlas}. In these cohorts, 3D structural T1-weighted (T1w) MRI is often the only modality available.

Automating lesion segmentation directly from multi-site T1w scans is therefore essential to unlock legacy datasets for retrospective research. Prior work on T1w-only stroke segmentation has evolved from classical feature-engineering approaches that plateau around Dice 0.40--0.50 \cite{ito2019comparison} to modern deep learning networks, which typically report Dice around 0.60 on multi-center T1w benchmarks \cite{liew2022atlas}. However, T1w lesion segmentation presents two major technical hurdles:
\begin{enumerate}
    \item \textbf{Intensity Ambiguity:} On single-channel T1w scans, ischemic lesions present as subtle hypointensities and tissue voids that blend into surrounding cerebrospinal fluid \cite{ito2019comparison}.
    \item \textbf{Multi-Center Domain Shift:} Scans acquired across disparate clinical centers vary in field strength (1.5T vs. 3T), head coil configurations, and pulse sequence parameters, creating substantial inter-site intensity variation \cite{liew2022atlas,petzsche2022isles}.
\end{enumerate}
Despite advances in 3D deep learning architectures, existing models struggle with acute lesions on unbalanced multi-center T1w benchmarks. The root cause is sample scarcity; acute cases are substantially underrepresented in the training distribution (see Section~\ref{sec:dataset} for cohort statistics).

To overcome this plateau, standard preprocessing and architectural modifications are commonly attempted. We evaluated several conventional remedies and found that N4 bias field correction, $Z$-axis spatial coordinate channels, and FiLM metadata conditioning all failed to improve performance. Detailed ablation results are reported in Section~\ref{sec:ablation}. These results indicate that correcting scanner-induced intensity variation and providing explicit spatial coordinates does not close the gap — the bottleneck is sample scarcity, not preprocessing.

Because acute sample scarcity remains the primary bottleneck in native T1w stroke segmentation, we introduce an on-the-fly 3D CarveMix data augmentation pipeline \cite{zhang2021carvemix} integrated into nnU-Net v2. The pipeline extracts real 3D lesion patches from donor cases in the training split and pastes them into recipient training patches using distance-transform edge blending.

By executing synthetic lesion placement dynamically during data loading, the network encounters continuous spatial and morphological lesion variations across training epochs. This dynamic formulation generates new placements without storing synthetic volumes on disk.

Our main contributions are:
\begin{itemize}
    \item We demonstrate empirically that global intensity preprocessing (N4) and spatial coordinate channels ($Z$-grid) do not resolve the performance bottleneck on native T1w stroke scans.
    \item We implement an on-the-fly 3D CarveMix augmentation engine within nnU-Net v2 that dynamically pastes real lesions using donors and recipients from the training split.
    \item We evaluate a \texttt{MedNeXt-L} ($k=5$) backbone for lesion segmentation on scans acquired in native space.
\end{itemize}

\section{Methods}

\subsection{Native Input Pipeline}
We use native-space T1w scans without atlas registration. nnU-Net resamples the volumes to a 1\,mm isotropic grid for training. Each volume is Z-score normalized based on the foreground brain voxels (computed per-volume). We experimented with N4 bias field correction \cite{tustison2010n4} and found it made no meaningful difference, so we skip it to keep the pipeline simple.

\subsection{Architecture: MedNeXt-L}
We use MedNeXt \cite{roy2023mednext}, a ConvNeXt-style 3D architecture that replaces standard convolutions with depthwise separable ones and adds inverted bottleneck blocks. We selected the large variant (MedNeXt-L) with kernel size $k=5$ based on its 5-fold cross-validation Dice at 1{,}000 epochs (0.663), where it outperformed the default 3D U-Net (0.657) and STU-Net-L (0.662) with the lowest cross-fold variance. The $5^3$ kernel provides a wider effective receptive field than the standard $3^3$ kernel, which helps distinguish subtle lesions from adjacent cerebrospinal fluid.

\subsection{Dynamic 3D CarveMix Data Augmentation}
The core idea of CarveMix \cite{zhang2021carvemix} is to take a lesion from one subject and paste it into another, creating a synthetic example with a known label. We implement this on the fly inside the nnU-Net training loop.

At startup, we extract up to 500 lesion patches from preprocessed training cases in the current fold. For each training patch, the transform attempts a paste with probability $p = 0.4$. It samples a donor with inverse-chronicity-frequency weighting, yielding approximately equal donor selection probabilities across the three chronicity groups. It selects a candidate paste center from voxels with segmentation label 0. The alpha mask is computed by dividing the donor lesion mask distance transform by 3.0 voxels and clipping to $[0,1]$. The segmentation label is updated as the union of the original and pasted lesion masks.

Pasting takes place during training data loading, so repeated samples from the same subject can receive different lesion placements. Within each fold, donor and recipient subjects come from the training split.

\section{Experimental Setup}

\subsection{Dataset and Validation}
\label{sec:dataset}
We use the ISLES 2026 benchmark dataset \cite{isles2026challenge}, built from the ATLAS stroke neuroimaging dataset \cite{liew2022atlas,atlas2026}: 1,453 native T1w scans from 55 clinical centers, each with a manually traced lesion mask. By chronicity, 92 subjects are acute ($\le 7$ days post-onset, 6.3\%), 712 are subacute (8--180 days, 49.0\%), and 649 are chronic ($>$180 days, 44.7\%), with a mean of 690 days post-stroke. The distribution is heavily skewed toward chronic and subacute cases, while acute cases constitute a small minority.

We used five-fold cross-validation with disjoint training and validation subjects within each fold. Acquisition centers could contribute subjects to both sets. The reported results therefore estimate performance on held-out subjects within a multi-center dataset.

\subsection{Training Implementation}
Training runs on full NVIDIA H100 GPUs using nnU-Net v2 \cite{isensee2021nnu} with the MedNeXt-L trainer. We use the default Dice + Cross-Entropy loss, SGD with Nesterov momentum ($\mu = 0.99$), initial learning rate 0.01 with polynomial decay, batch size 2, and patch size $128 \times 128 \times 128$. MedNeXt-L with CarveMix was trained for 500 epochs. The plain MedNeXt-L reference was trained for 1{,}000 epochs. Its estimated performance at epoch 500 was obtained by interpolation from its training trajectory.

We fix the number of iterations per epoch to 250, regardless of dataset size. With 1,453 subjects, each subject is sampled roughly once every 6 epochs on average; the network sees a different subset of the training data in each pass. The CarveMix augmentation further increases diversity within those 250 iterations by synthesizing new lesion placements on the fly.

\section{Results and Ablation}

\subsection{Baseline Performance}
To quantify the impact of sample imbalance, we evaluate a MedNeXt-L baseline stratified by stroke stage. At 1{,}000 epochs the model achieves $DSC = 0.407$ on acute cases, $DSC = 0.666$ on subacute cases, and $DSC = 0.696$ on chronic cases, with the lowest Dice in the acute group. At epoch 500, the interpolated baseline estimates are approximately $0.374$ for acute, $0.633$ for subacute, and $0.663$ for chronic cases.

The 500-epoch CarveMix run achieved a mean Dice of 0.648, which was 0.018 higher than the interpolated plain MedNeXt-L estimate of 0.630. The measured CarveMix scores also exceeded the corresponding interpolated estimates in each chronicity group (Table~\ref{tab:chronicity}). A separately trained 500-epoch plain MedNeXt-L control would be needed to estimate the effect of CarveMix under the same training schedule.

\begin{table}[t]
\centering
\caption{Chronicity-stratified 5-fold cross-validation Dice. Baseline values at 500 epochs are estimates interpolated from the 1{,}000-epoch training trajectory.}
\label{tab:chronicity}
\begin{tabular}{l c c c}
\toprule
\textbf{Stage} & \textbf{MedNeXt-L (500e est.)} & \textbf{+ CarveMix} & \textbf{$\Delta$} \\
\midrule
Acute ($\le 7$d) & 0.374 & 0.393 & +0.019 \\
Subacute (8--180d) & 0.633 & 0.651 & +0.018 \\
Chronic ($>180$d) & 0.663 & 0.681 & +0.018 \\
\midrule
Overall mean & 0.630 & 0.648 & +0.018 \\
\bottomrule
\end{tabular}
\end{table}

\subsection{Ablation of Preprocessing and Augmentation}
\label{sec:ablation}
Table~\ref{tab:ablation} summarizes the comparisons with the corresponding baselines. N4 bias field correction \cite{tustison2010n4}, applied to a ResEnc-M baseline (0.639), yielded a marginal gain of $\Delta$Dice = +0.002. Appending a relative $Z$-axis spatial coordinate as a second input channel to provide explicit height context lowered MedNeXt-S performance from 0.636 to 0.616 ($\Delta = -0.020$), indicating the network does not benefit from explicit spatial coordinates. FiLM metadata conditioning \cite{perez2018film} on ResEnc-XL similarly degraded the result from 0.656 to 0.649 ($\Delta = -0.007$).

The CarveMix row compares the measured 500-epoch score of 0.648 with the interpolated plain MedNeXt-L estimate of 0.630 (Table~\ref{tab:ablation}).

\begin{table}[t]
\centering
\caption{Ablation study showing 5-fold cross-validation Dice on native T1w.}
\label{tab:ablation}
\begin{tabular}{l c c c}
\toprule
\textbf{Configuration} & \textbf{Dice $\uparrow$} & \textbf{$\Delta$Dice} & \textbf{Std. Dev.} \\
\midrule
  + N4 Correction (vs.\ ResEnc-M 0.639) & 0.641 & +0.002 & $\pm$0.005 \\
  + $Z$-Grid Channel (vs.\ MedNeXt-S 0.636) & 0.616 & $-$0.020 & $\pm$0.033 \\
  + FiLM Conditioning (vs.\ ResEnc-XL 0.656) & 0.649 & $-$0.007 & $\pm$0.010 \\
\midrule
  \textbf{+ CarveMix} (vs.\ MedNeXt-L 500e est.\ 0.630) & \textbf{0.648} & \textbf{+0.018} & $\pm$0.019 \\
\bottomrule
\end{tabular}
\end{table}

\section{Conclusion}

The performance ceiling on native T1w stroke segmentation (Dice $\approx 0.66$) is not caused by bias field inhomogeneity or missing global spatial context. N4 correction and $Z$-height coordinate channels produce negligible or negative improvement; the $Z$-grid lowered MedNeXt-S Dice from 0.636 to 0.616 ($\Delta = -0.020$). The bottleneck is acute lesion scarcity: 92 cases out of 1,453, with minimal tissue contrast in the early post-stroke period. Dynamic 3D CarveMix with MedNeXt-L achieved a mean five-fold Dice of 0.648 on held-out subjects. Its score exceeded the interpolated 500-epoch plain MedNeXt-L estimate by 0.018. A directly matched training comparison is needed to determine how much of this difference is attributable to the augmentation. We note that with 92 acute subjects, per-stage metrics are inherently noisy, and further validation on prospectively collected acute cohorts is warranted.

\bibliographystyle{splncs04}
\bibliography{ref}

@article{feigin2021gbd,
  author  = {Feigin, Valery L. and Stark, Benjamin A. and Johnson, Catherine Owens and Roth, Gregory A. and Bisignano, Catherine and Abady, Gdiom Gebreheat and Abbasifard, Mitra and Abbasi-Kangevari, Mohsen and others},
  title   = {Global, regional, and national burden of stroke and its risk factors, 1990--2019: a systematic analysis for the {Global Burden of Disease Study} 2019},
  journal = {The Lancet Neurology},
  volume  = {20},
  number  = {10},
  pages   = {795--820},
  year    = {2021},
}

@article{petzsche2022isles,
  author  = {Hernandez Petzsche, Moritz R. and de la Rosa, Ezequiel and Hanning, Uta and Wiest, Roland and Valenzuela, Waldo and Reyes, Mauricio and Meyer, Maria I. and Liew, Sook-Lei and Kofler, Florian and Kirschke, Jan S. and Wiestler, Benedikt},
  title   = {{ISLES} 2022: A multi-center magnetic resonance imaging stroke lesion segmentation dataset},
  journal = {Scientific Data},
  volume  = {9},
  number  = {1},
  pages   = {762},
  year    = {2022}
}

@inproceedings{perez2018film,
  author  = {Perez, Ethan and Strub, Florian and de Vries, Harm and Dumoulin, Vincent and Courville, Aaron},
  title     = {{FiLM}: Visual Reasoning with a General Conditioning Layer},
  booktitle = {Proceedings of the AAAI Conference on Artificial Intelligence},
  volume    = {32},
  number    = {1},
  pages     = {3942--3951},
  year      = {2018}
}

@article{liew2022atlas,
  author  = {Liew, Sook-Lei and Lo, Bethany P. and Donnelly, Miranda R. and Zavaliangos-Petropulu, Artemis and Jeong, Jessica N. and Barisano, Giuseppe and Hutton, Alexandre and Simon, Julia P. and Juliano, Julia M. and Suri, Anisha and others},
  title   = {A large, curated, open-source stroke neuroimaging dataset to improve lesion segmentation algorithms},
  journal = {Scientific Data},
  volume  = {9},
  number  = {1},
  pages   = {320},
  year    = {2022}
}

@misc{atlas2026,
  author       = {{International Neuroimaging Data-sharing Initiative}},
  title        = {Anatomical Tracings of Lesion After Stroke ({ATLAS})},
  howpublished = {\url{https://fcon_1000.projects.nitrc.org/indi/retro/atlas.html}},
  year         = {2026},
  note         = {Accessed 2026}
}

@article{isensee2021nnu,
  author  = {Isensee, Fabian and Jaeger, Paul F. and Kohl, Simon A. A. and Petersen, Jens and Maier-Hein, Klaus H.},
  title   = {{nnU-Net}: a self-configuring method for deep learning-based biomedical image segmentation},
  journal = {Nature Methods},
  volume  = {18},
  number  = {2},
  pages   = {203--211},
  year    = {2021}
}

@inproceedings{roy2023mednext,
  author  = {Roy, Saikat and Koehler, Gregor and Ulrich, Constantin and Baumgartner, Michael and Petersen, Jens and Isensee, Fabian and Jaeger, Paul F. and Maier-Hein, Klaus H.},
  title     = {{MedNeXt}: Transformer-Driven Scaling of {ConvNets} for Medical Image Segmentation},
  booktitle = {Medical Image Computing and Computer Assisted Intervention -- MICCAI 2023},
  pages     = {405--415},
  year      = {2023},
  publisher = {Springer}
}

@article{zhang2021carvemix,
  author  = {Zhang, Xinru and Liu, Chenghao and Ou, Ni and Zeng, Xiangzhu and Zhuo, Zhizheng and Duan, Yunyun and Xiong, Xiaoliang and Yu, Yizhou and Liu, Zhiwen and Liu, Yaou and Ye, Chuyang},
  title   = {{CarveMix}: A Simple Data Augmentation Method for Brain Lesion Segmentation},
  journal = {NeuroImage},
  volume  = {271},
  pages   = {120041},
  year    = {2023},
  note    = {Earlier conference version in MICCAI 2021, pp. 196--205}
}

@article{tustison2010n4,
  author  = {Tustison, Nicholas J. and Avants, Brian B. and Cook, Philip A. and Zheng, Yuanjie and Egan, Alexander and Yushkevich, Paul A. and Gee, James C.},
  title   = {{N4ITK}: Improved {N3} Bias Correction},
  journal = {IEEE Transactions on Medical Imaging},
  volume  = {29},
  number  = {6},
  pages   = {1310--1320},
  year    = {2010}
}

@article{ito2019comparison,
  author  = {Ito, Kaori L. and Kim, Hosung and Liew, Sook-Lei},
  title   = {A comparison of automated lesion segmentation approaches for chronic stroke {T1}-weighted {MRI} data},
  journal = {Human Brain Mapping},
  volume  = {40},
  number  = {16},
  pages   = {4669--4685},
  year    = {2019}
}

@misc{isles2026challenge,
  author       = {{ISLES 2026 Organizing Committee}},
  title        = {Ischemic Stroke Lesion Segmentation Challenge ({ISLES}) 2026},
  howpublished = {Zenodo},
  year         = {2026},
  doi          = {10.5281/zenodo.19856506},
}

\end{document}